\documentclass[%
reprint,pra,
superscriptaddress,
amsmath,amssymb,
aps,
]{revtex4}

\usepackage{graphicx}% Include figure files
\usepackage{dcolumn}% Align table columns on decimal point
\usepackage{bm}% bold math
\usepackage{hyperref}% add hypertext capabilities
\usepackage[utf8]{inputenc}
\usepackage{subfigure}
\usepackage{cleveref}
\usepackage{amsmath}
\begin{document}
	
\preprint{APS/123-QED}

\title{0.8\% Nyquist computational ghost imaging via non-experimental deep learning}

\author{Haotian Song}
\affiliation{School of Physics, Xi'an Jiaotong University, Xi'an, Shaanxi 710049, China}
\affiliation{College of Physics \& Astronomy, University of Manchester, Manchester M13 9PL, UK}

\author{Xiaoyu Nie}
\affiliation{%
	School of Physics, Xi'an Jiaotong University, Xi'an, Shaanxi 710049, China}%
\affiliation{%
	Texas A\&M University, College Station, Texas, 77843, USA}%

\author{Hairong Su}
\affiliation{School of Mathematics and Statistics, Xi'an Jiaotong University, Xi'an, Shaanxi 710049, China}

\author{Hui Chen}
\affiliation{Electronic Materials Research Laboratory, Key Laboratory of the Ministry of Education \& International Center for Dielectric Research, Xi’an Jiaotong University, Xi’an, 710049, China}

\author{Yu Zhou}
\affiliation{School of Physics, Xi'an Jiaotong University, Xi'an, Shaanxi 710049, China}

\author{Xingchen Zhao}%
\affiliation{%
	Texas A\&M University, College Station, Texas, 77843, USA}%
\author{Tao Peng}%
\email{taopeng@tamu.edu}
\affiliation{%
	Texas A\&M University, College Station, Texas, 77843, USA}%
\author{Marlan O. Scully}%
\affiliation{%
	Texas A\&M University, College Station, Texas, 77843, USA}%
\affiliation{%
Baylor University, Waco, 76706, USA}%
\affiliation{%
Princeton University, Princeton, NJ 08544, USA}%
\date{\today}

\begin{abstract}
We present a framework for computational ghost imaging based on deep learning and customized pink noise speckle patterns. The deep neural network in this work, which can learn the sensing model and enhance image reconstruction quality, is trained merely by simulation. To demonstrate the sub-Nyquist level in our work, the conventional computational ghost imaging results, reconstructed imaging results using white noise and pink noise via deep learning are compared under multiple sampling rates at different noise conditions. We show that the proposed scheme can provide high-quality images with a sampling rate of 0.8\% even when the object is outside the training dataset, and it is robust to noisy environments.  This method is excellent for various applications, particularly those that require a low sampling rate, fast reconstruction efficiency, or experience strong noise interference.
\end{abstract}

	\maketitle
	
\section{Introduction}
Ghost imaging (GI)~\cite{Pittman1995Optical,bennink2002two,valencia2005two,chen2009lensless} is an innovative method for measuring the spatial correlations between light beams. With GI, the signal light field interacts with the object and is collected by a single-pixel detector, and the reference light field, which does not interact with the object, falls onto the imaging detector. Therefore, the image information is not present in either beam alone but only revealed in their correlations. Computational ghost imaging (CGI)~\cite{bromberg2009ghost,shapiro2008computational} was proposed to further ameliorate and simplify this framework. In CGI, The reference arm that records the speckles is replaced by loading pre-generated patterns directly onto the spatial light modulator or the digital micromirror device (DMD). The unconventional image is then revealed by correlating the sequentially recorded intensities at the single-pixel detector with the corresponding patterns. CGI finds a lot of applications such as wide spectrum imaging \cite{shrekenhamer2013terahertz,aspden2015photon,klein2019x}, remote sensing~\cite{hardy2013computational}, and quantum-secured imaging~\cite{clemente2010optical}.

However, CGI generally requires a large number of samplings to reconstruct a high-quality image, or the signal would have been submerged under correlation fluctuations and environmental noise. To suppress the environmental noise and correlation fluctuations, the required minimum number of sampling is proportional to the total pixel number of the pattern applied on DMD, \textit{i.e.}, the Nyquist sampling limit~\cite{cook1986stochastic,tropp2009beyond}. The image could have a meager quality with a limited sampling number. This demanding requirement hindered CGI from fully replacing conventional photography. A large number of schemes have been proposed to improve CGI’s speed and decrease the sampling rate (sub-Nyquist). For instance, compressive sensing imaging can reconstruct images with a relatively low sampling rate by exploiting the sparsity of the objects~\cite{Magana2013Compressive,Katz2009Compressive,Xu:18,Yi2019Compressive}. It nevertheless largely depends on the sparsity of objects and is sensitive to noise~\cite{du2012influence}. Orthonormalized noise patterns can be used to suppress the noise and improve the image's quality under a limited sampling number \cite{luo2018orthonormalization,nie2020sub}. In particular, the orthonormalized colored noise patterns can break the Nyquist limit down to $\sim 5\%$~\cite{nie2020sub}. Fourier and sequence-ordered Walsh-Hadamard patterns, which are orthogonal to each other in time or spatial domain, were also applied to the sub-Nyquist imaging~\cite{zhang2015single,wang2016fast,Zhang:17}. The Russian doll~\cite{sun2017russian} and cake-cutting~\cite{yu2019super} ordering of Walsh-Hadamard patterns can minimize the sampling ratio to 5\%-10\% Nyquist limit. 

Recently, the deep learning (DL) technique is employed to identify images~\cite{he2020handwritten,li2020object} and improve the quality of images with the deep neural network (DNN)~\cite{Lyu2017Deep,He2018Ghost,shimobaba2018computational,wang2019learning,zhai2019foveated,wu2020sub,rizvi2020deepghost,wu2020deep,bian2020residual}. Specifically, computational ghost imaging via deep learning (CGIDL) has shown a minimum ratio of Nyquist limit down to $\sim 5\%$ \cite{He2018Ghost,wu2020sub}. However, such work's DNNs are trained by experimental CGI results. Only when the training environment is highly identical to the environment used for image reconstruction can the DNN be effective. This limits its universal applications and restricts it to achieve quick reconstructions. Usually at least thousands of inputs have to be generated for the training, which would be very time-consuming if conducting experimental training each time. Some studies have been performed to test the effectiveness of non-experimental CGI training DNN, the minimum ratios of the Nyquist limit were up to a few percent~\cite{shimobaba2018computational,wang2019learning, wu2020deep}. However, the sampling ratio is much higher for objects outside of training dataset than those in the training dataset~\cite{wu2020sub}. Therefore, despite the proliferation of numerous algorithms, retrieving high-quality images outside of the training group with a meager Nyquist limit ratio by non-experimental training remains a challenge for the CGIDL system.

This letter aims to minimize the necessary sampling number further and improve the imaging quality with the combination of DL and colored noise CGI. Recently, it has been shown that the synthesized colored noise patterns possess unique non-zero correlations between neighborhood pixels via amplitude modulation in the spatial frequency domain~\cite{nie2021noise,li2021sub}. In particular, The pink noise CGI owns positive cross-correlations in the second-order correlation~\cite{nie2021noise}. It gives a good image quality under a boisterous environment or pattern distortion when the traditional CGI method fails. Combining DL with pink noise CGI shows that the imaging can be retrieved under an extremely low sampling rate ($\sim 0.8\%$). We also show that we can get training patterns from the simulation without introducing the environmental noises, \textit{i.e.}, there is no need to get DNN training with a large number of experimental training inputs. The object used in the experiment can be independent of the training dataset, which can largely benefit CGIDL in the real application.

\section{deep learning}
\subsection{Deep neural network}
\begin{figure}[!ht]
    \centering\includegraphics[width=0.8\linewidth]{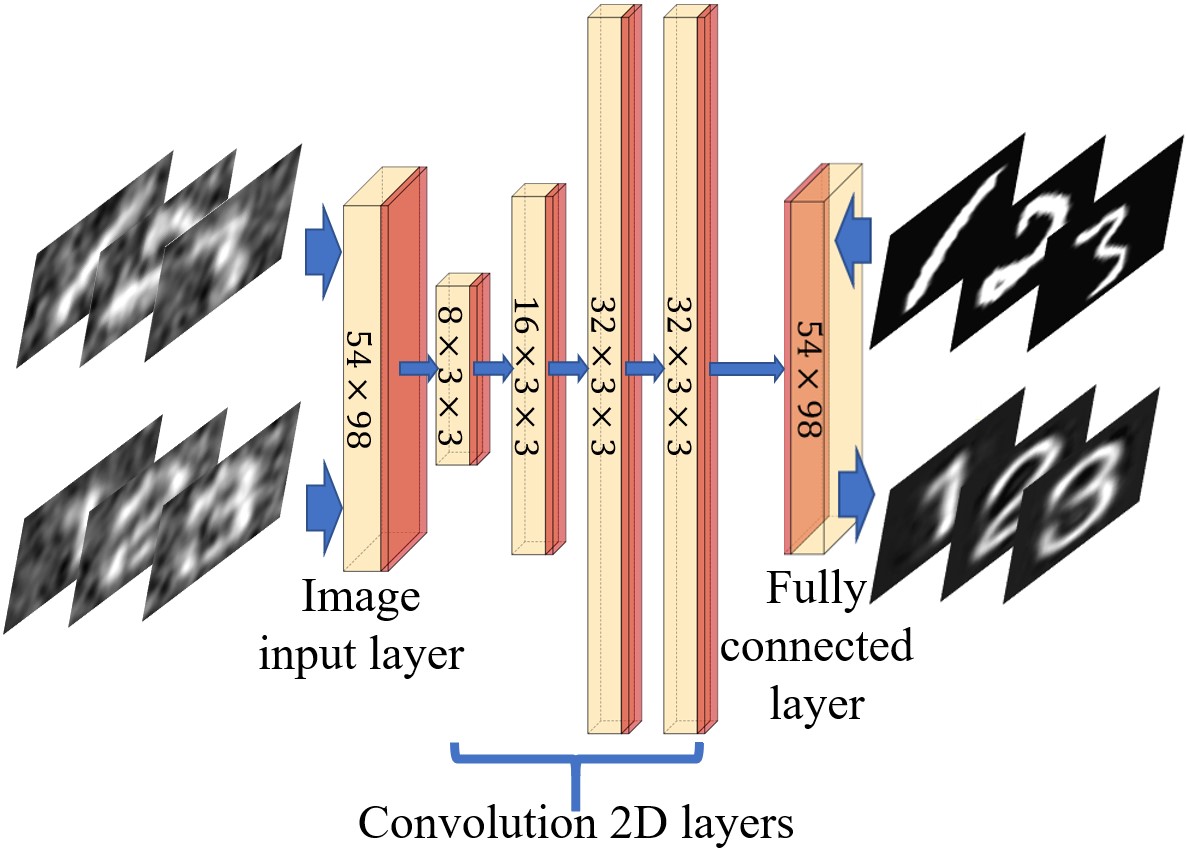}
    \caption{Architecture of DNN. It consists of four convolution layers, one image input layer, one fully connected layer (yellow), the rectified linear unit, and the batch normalization layers (red). In the upper line are CGI results (training inputs) and handwriting ground truths (training labels); In the bottom line are CGI results from the experiment (test inputs) and CGIDL results (test outputs) with block style.}
    \label{fig:DNN}
\end{figure}
Our DNN model, as shown in Fig.~\ref{fig:DNN}, uses four convolution layers, one image input layer, and one fully connected layer. Small $3\times3$ receptive fields are applied throughout the whole convolution layers for better performance~\cite{szegedy2015going}. Batch normalization layers (BNL), rectified Linear Unit (ReLU) layers and zero padding are added between each convolution layer. The BNL is functioned to avoid internal covariate shift during the training process and speed up the training of DNN~\cite{ioffe2015batch}. The ReLU layer applies a threshold operation to each element of the inputs~\cite{10.5555/3104322.3104425}. The zero padding part was designed to maintain the characteristic of input images' boundaries. To customize the size of training pictures, both the input and output layers are set to be $54\times98$. The solver for training is employed by the Stochastic Gradient Descent with Momentum Optimizer (SGDMO) to reduce the oscillation via using momentum. The parameter vector can be updated via equation Eq.~(\ref{eq:1}), which demonstrates the updating process during the iteration.
\begin{equation}
    \label{eq:1}
    \theta_{\ell+1}=\theta_{\ell}-\alpha \nabla E\left(\theta_{\ell}\right)+\gamma\left(\theta_{\ell}-\theta_{\ell-1}\right),
\end{equation}
where $\ell$ is the iteration number, $\alpha$ is the learning rate, $\theta$ is the parameter vector, and $E(\theta)$ is the loss function, mean square error (MSE). The MSE is defined as
\begin{equation}
\mathrm{MSE} = \frac{1}{N_{\mathrm{pixel}}}\sum_{i=1}^{N_{\mathrm{pixel}}}{[\frac{G_i-X_i}{\langle G_{(o)}\rangle }]^2}.
\label{eq:mse}
\end{equation}

Here, $G$ represents the pixel value of the resulted imaging. $G_{(o)}$ represents pixels that the light ought to be transmitted, \textit{i.e.}, the object area, while $G_{(b)}$ represents pixels that the light ought to be blocked, \textit{i.e.}, the background area. $X$ is the ground truth calculated by
\begin{equation}
{X_i} =  
\begin{cases}
\langle G_{(o)}\rangle, & \text{ Transmission = 1}\\
\langle G_{(b)}\rangle, & \text{ Transmission = 0}
\end{cases}
\end{equation}

The third part on the right hand side of the equation is the feature of SGDMO, analog to the momentum where $\gamma$ determines the contribution of the previous gradient step to the current iteration~\cite{murphy2012machine}. Two strategies are applied to avoid over-fitting of training images. At the end of DNN, a dropout layer is applied with probability of dropping out input elements being $0.2$, which is aimed to reduce the connection between convolution layers and the fully connected layer~\cite{JMLR:v15:srivastava14a}. Meanwhile, we adopted a step decay schedule for the learning rate. The learning rate dropped from $10^{-3}$ to $10^{-4}$ after 75 epochs, which constrain the fitting parameters within a reasonable region. Lower the learning rate could avoid overfitting significantly with constant maximum epochs.

\subsection{Network training}
\begin{figure*}[!htbp]
    \centering
    \includegraphics[width=0.85\linewidth]{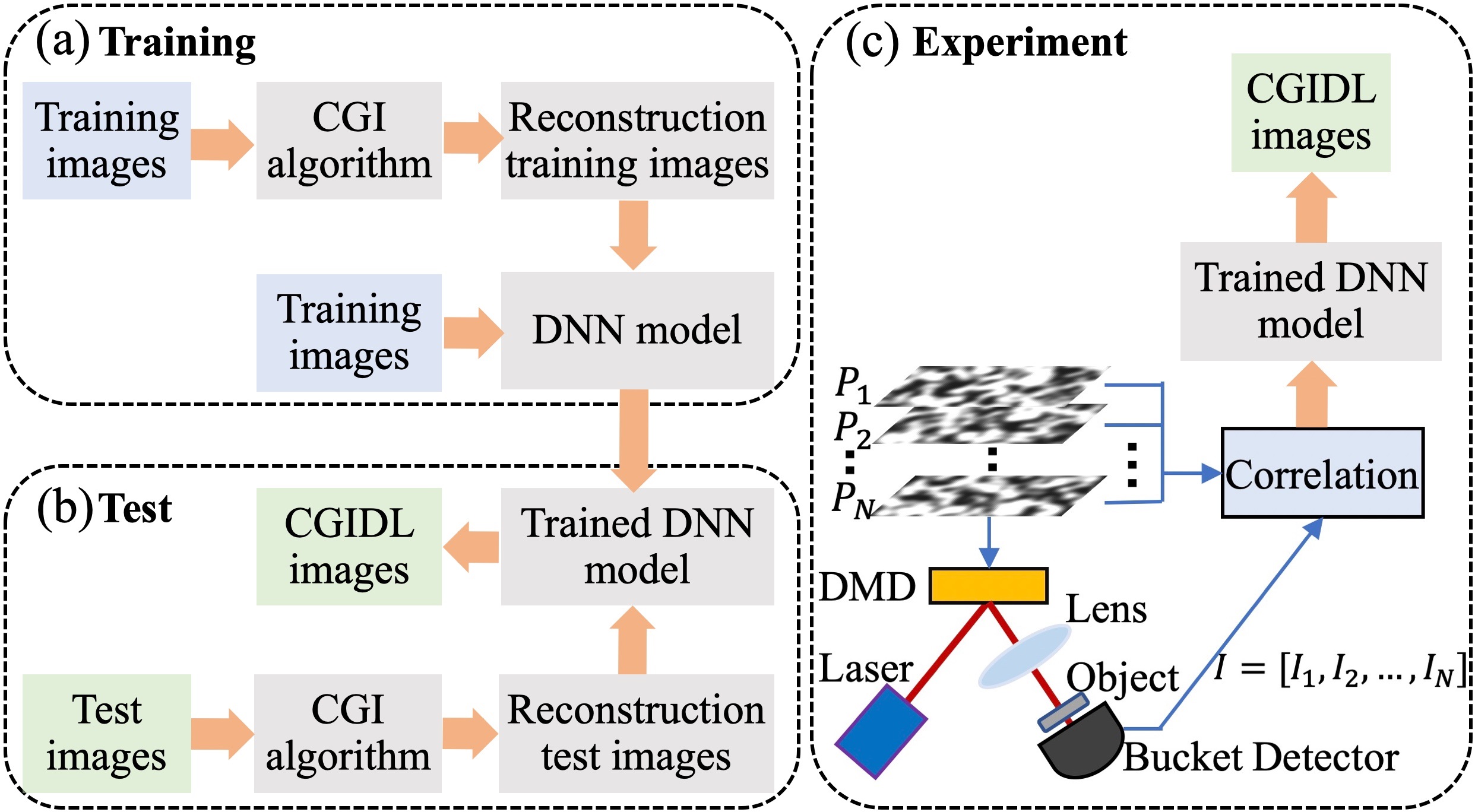}
    \caption{The flow chart of CGIDL consists of three parts: (a) training, (b) test, and (c) experiment. The DNN model is trained with CGI results from database via simulation. The simulation testing process and experimental measuments use both the handwriting digits and block style digits. The experimental part for CGI uses pink noise and white noise speckle patterns, and their CGI results are ameliorated by trained DNN model.}
    \label{fig:flowchart}
\end{figure*}

The proposed CGIDL scheme requires a training process based on pre-prepared dataset. After training in simulation, it owns ability to reconstruct the images. We use a set of 10000 digits from the MNIST handwritten digit database~\cite{deng2012mnist} as training images. All images are resized and normalized to $54\times98$ to test the smaller sampling ratio. These training images are reconstructed by the CGI algorithm. The training images and reconstruction training images then feed the DNN model as inputs and outputs, respectively, as shown in Fig.~\ref{fig:flowchart}(a). The white noise and pink noise speckle patterns are used separately for the training process, using exactly the same protocol.  The maximum epochs are set as 100, and the training iteration is 31200. The program is implemented via MATLAB R2019a Update 5 (9.6.0.1174912, 64-bit), and the DNN is implemented through DL Toolbox. The GPU-chip NVIDIA GTX1050 is used to accelerate the speed of the computation.

The trained DNN is then tested by simulation and used for retrieving CGI results in the experiments. In the testing part, the CGI algorithm generates reconstructed images from testing images with both the MNIST handwritten digits and block style digits, where the later set is different from images in the training group. As shown in Fig.~\ref{fig:flowchart}(b), the trained DNN, fed with reconstruction testing images, generates CGIDL results. Comparing the difference between CGIDL and testing images, we could measure the quality of the trained DNN. Well-performed DNN can be used for retrieving CGI in the experiment.

The schematic of the experiment is shown in Fig.~\ref{fig:flowchart}(c). A CW laser is used to illuminate the DMD, on which the noise patterns are loaded. The pattern generated by the DMD is then projected onto the object. In our experiment, the size of the noise patterns is $216 \times 392$ DMD pixels ($54 \times 98$ independent pixels), in which the independent changeable mirrors count for $4 \times 4$ pixels. Each DMD pixel is $ 16\mu m \times 16 \mu m$ in size.  

In the CGI process, the quality of the images is proportional to the sampling rate, which is the ratio between the number of illumination patterns $N_{\mathrm{pattern}}$ and $N_{\mathrm{pixel}}$~\cite{ferri2010differential, wang2015gerchberg}:
\begin{equation}
    \beta=N_{\mathrm{pattern}}/N_{\mathrm{pixel}}.
\end{equation}

In the following, We compared the trained network using white noise speckle patterns (DL white) and pink noise speckle patterns (DL pink), as well as the conventional CGI (CGI white) in terms of reconstruction performance with
respect to the sampling ratio $\beta$.

\section{Simulation}

\begin{figure}[!ht]
  \centering
  \includegraphics[width=0.85\linewidth]{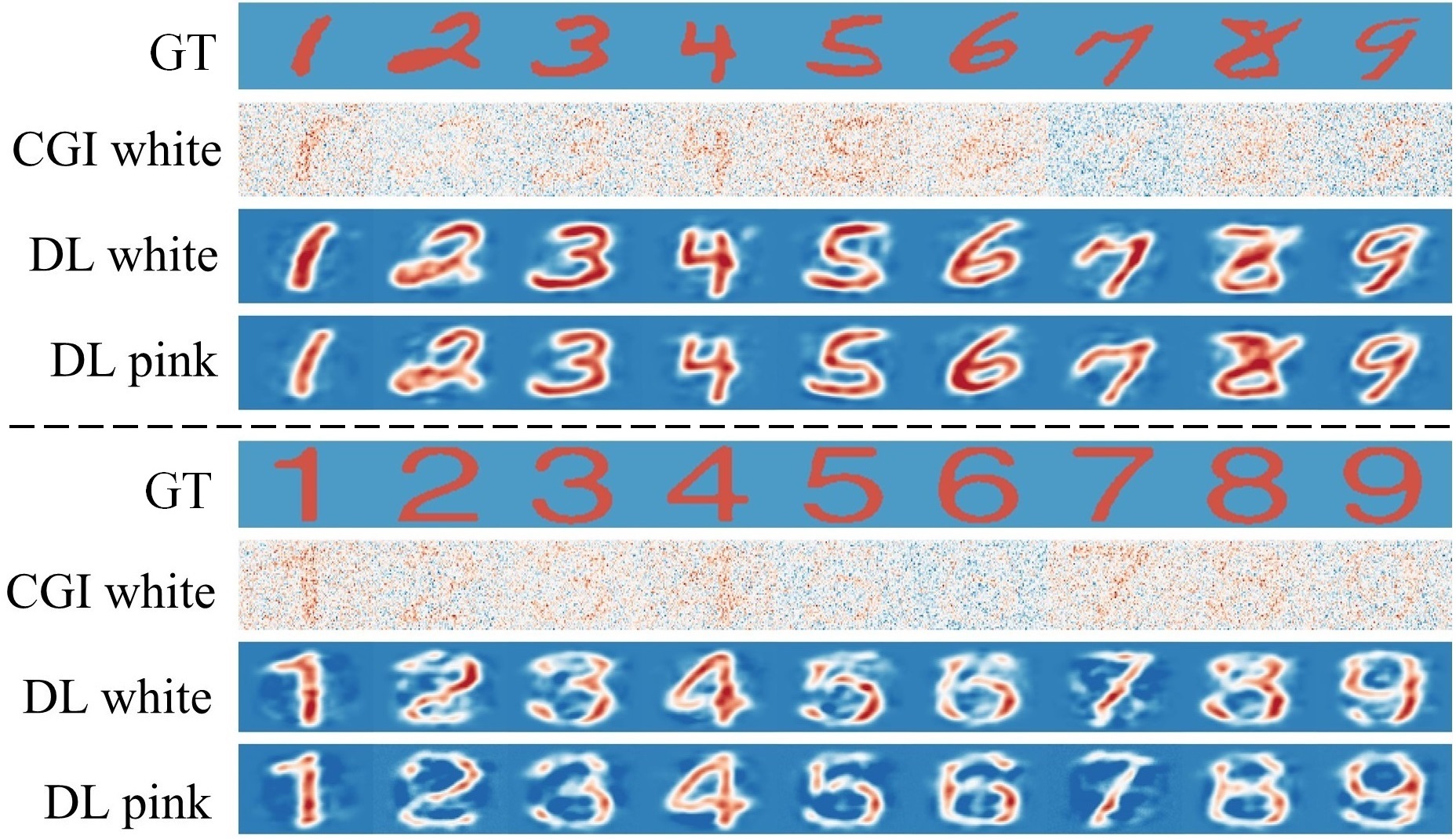}
    \caption{Simulation results without noise. The upper part used handwriting digits 1-9 from the training dataset, and the lower part used block style digits 1-9, which is outside the training dataset. All the simulations are done at $\beta =5\%$. GT: ground truth.}
    \label{fig:simu_nonoise}
\end{figure}

To test the robustness of our method to different datasets, noise, and its performs at different sampling rates, we performed a set of simulations.  Two sets of testing images are used in the simulation. One of which is the handwriting digits 1-9 from  the training set, the other is the block style digits 1-9, which are completely independent of training images.  These images have $28\times28$ pixels and are resized into $54\times98$ by widening and amplification. We started our simulation from the comparison of the CGI white, DL white and DL pink without noise at $\beta=5\%$, as shown in Fig.~\ref{fig:simu_nonoise}. The upper part is with the handwriting digits 1-9, the lower part is with the block style digits 1-9. Apparently, at this low sampling rate, the traditional CGI method fails to retrieve the images in both cases. On the other hand, both DL methods work much better than the traditional CGI. For digits from the training dataset, both methods work almost equally well. For digits from outside the training dataset, DL pink works already better than DL white. For example, the DL white barely can distinguish digits '3' and '8', but DL pink can retrieve all the digits images. 

\begin{figure}[!ht]
    \centering\includegraphics[width=0.85\linewidth]{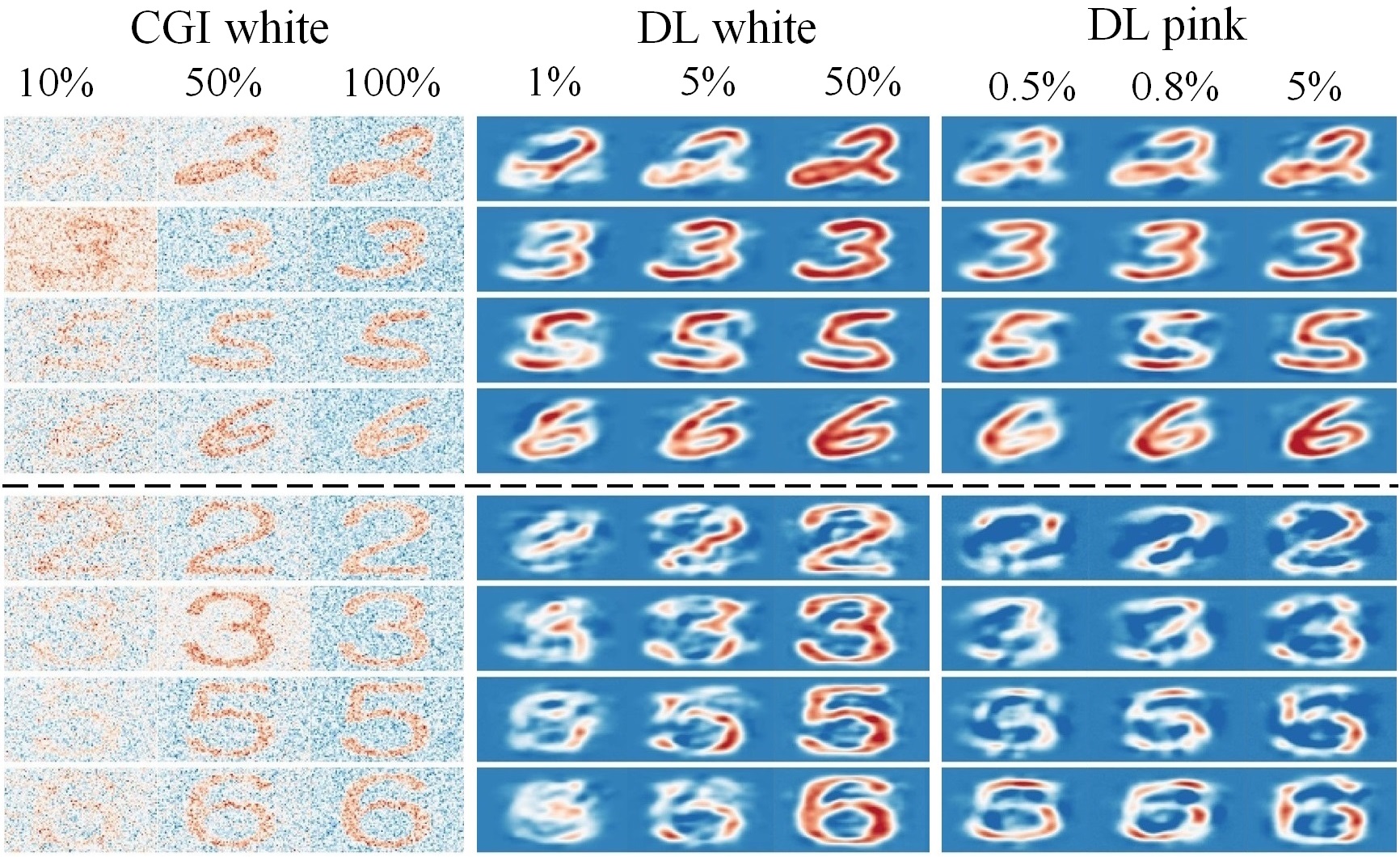}
    \caption{Simulation results of handwriting (top) and block style (bottom) digits 2, 3, 5, 6  with the SNR of $4.77~\mathrm{dB}$. The results of CGI white are done at $\beta$ of  10\%, 50\%, and 100\%, DL white with $\beta$ of 1\%, 5\%, and 50\%, and DL pink with $\beta$ of 0.5\%, 0.8\%, and 5\%.}
    \label{fig:simu_noise}
\end{figure}

In real application, there always exist noise in the measurement. It is therefore worthwhile checking the performances of different methods under the influence of noise. We then performed another set of simulations with added grayscale random noise. The signal-to-noise ratio (SNR) in logarithmic decibel scale is defined as
\begin{equation}
\mathrm{SNR} =10 \log\frac{P_{\mathrm{s}}}{P_{\mathrm{b}}},
\end{equation}
where  $P_{\mathrm{s}}$ is the average signal and $P_{\mathrm{b}}$ is the average noise background. Here the SNR is set to be $4.77 \mathrm{dB}$. 
As shown in Fig.~\ref{fig:simu_noise}, the upper part is the simulation with digits 2, 3, 5, and 6 from the training dataset, and the lower part is the simulation with digits 2, 3, 5, and 6 from the block style dataset. For both datasets, $\beta$ of 100\%, 50\%, and 10\% are chosen for CGI white, 50\%, 5\%, and 1\% for DL white, 5\%, 0.8\%, and 0.5\% for DL pink. The image quality is better with the increase of $\beta$ for all cases, as expected. As for the CGI white case, it can only give marginally visible images when the sampling rate is beyond $50\%$. The DL white, can retrieve the digits from the training dataset when $\beta=1\%$. However, for the block style digits, it fails to do so even when $\beta=5\%$. Unlike the previous case with no noise, there is a significant difference between objects from the training dataset and outside the training dataset. Lastly, we note that the DL pink trained network can retrieve the training dataset when $\beta=0.5\%$. It can also retrieve clear images for the block style digits at $\beta=0.8\%$. If we compare the black style images at $\beta=5\%$ for both DL white and DL pink with the no noise case in Fig.~\ref{fig:simu_nonoise}, it obvious that the quality of the former is largely affected by the noise, and the latter is barely affected. 

\section{Experiment}
\begin{figure}[!hpbt]
\centering\includegraphics[width=0.85\linewidth]{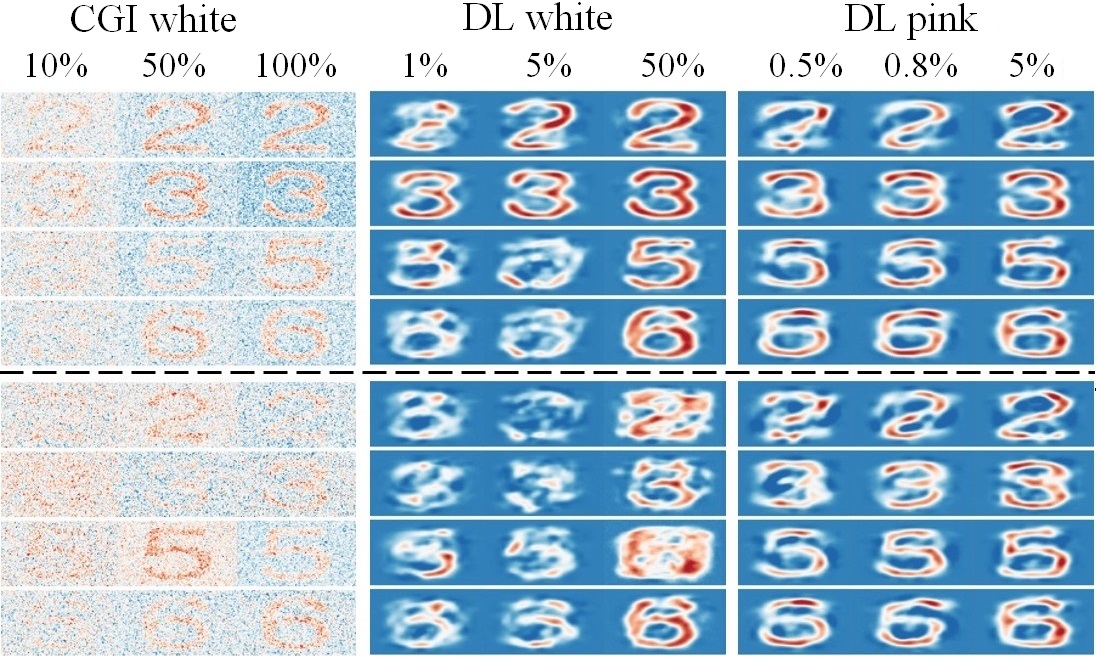}
\caption{Experimental results with the SNR of $14.90~\mathrm{dB}$ (upper) and  $4.77~\mathrm{dB}$ (lower). Objects are block style digits 2,3, 5, 6. Different sampling rates are shown for different methods: CGI white are done at $\beta$ of  10\%, 50\%, and 100\%, DL white with $\beta$ of 1\%, 5\%, and 50\%, and DL pink with $\beta$ of 0.5\%, 0.8\%, and 5\%.}
\label{fig:exp_result}
\end{figure}
To further demonstrate the advantage and applicability of CGIDL with pink noise, we perform experiments with the non-experimental and one-time trained model. All the experiments are done with digits 2, 3, 5, and 6 with block style. The block style is chosen to better compare the different behaviors of all three methods. 
We manage to start from a relatively low noise level of $\mathrm{SNR}=14.90~\mathrm{dB}$. The results are shown in upper part of Fig~\ref{fig:exp_result}. We can see at this noise level, the CGI white method barely can distinguish the images from the noisy background even at $\beta=100\%$. The DL white trained network, while giving clear images at $\beta=50\%$, fails to fully image the digits at $\beta=5\%$. This is mainly due to the objects are outside the training set, reveal one of the shortcomings of the standard DL network. On the other hand, our DL pink trained network can still give clear results even when $\beta=0.5\%$. 

We then increase the noise level to $\mathrm{SNR}=4.77~\mathrm{dB}$, which is the same as the simulation case so we can have a fair comparison. The experimental results are shown in the lower part of Fig~\ref{fig:exp_result}. The CGI white completely fail to image the digits even at $\beta=100\%$. The DL white trained network is also largely affected by the noisy environment, and not able to fully retrieve the images at $\beta=50\%$. On the other hand, the DL pink method can still image all digits at the sampling rate of $0.8\%$. If we compare these results to the corresponding low noise case, we can see that the image qualities do not change much, indicating our trained network is robust to noise. Also, the results with $\beta=0.8\%$ is better than the standard DL white network at $\beta=50\%$, which is about two orders higher. 

To quantitatively justify the quality of reconstructed block style images, we compare three evaluating indicators of image quality,  \textit{i.e.}, the peak signal to noise ratio (PSNR), the visibility (VIS), and the correlation coefficient (CC):
\begin{align}
\mathrm{PSNR}& = 10\times{\log_{10}[\frac{(2^k-1)^2}
	{\mathrm{MSE}}]},\cr
\mathrm{VIS} &= \frac{\langle G_{(\mathrm{o})}\rangle -\langle G_{(\mathrm{b})}\rangle}{\langle G_{(\mathrm{o})}\rangle+\langle G_{(\mathrm{b})}\rangle}, \cr
\mathrm{CC}& = \frac{\mathrm{Cov}(G,X)}{\sqrt{\mathrm{Var}(G)\mathrm{Var}(X)}}.
	\end{align}
Here $\mathrm{MSE}$ is defined in Eq.~\ref{eq:mse}, $\mathrm{Var(\,)}$ is the variance of its arguments, $\mathrm{Cov(\,)}$ is the covariance of its arguments, $k$ is the gray level of the image, and in our experiment $k\equiv 8$. 

\begin{figure}[!htp]
\centering\includegraphics[width=0.85\linewidth]{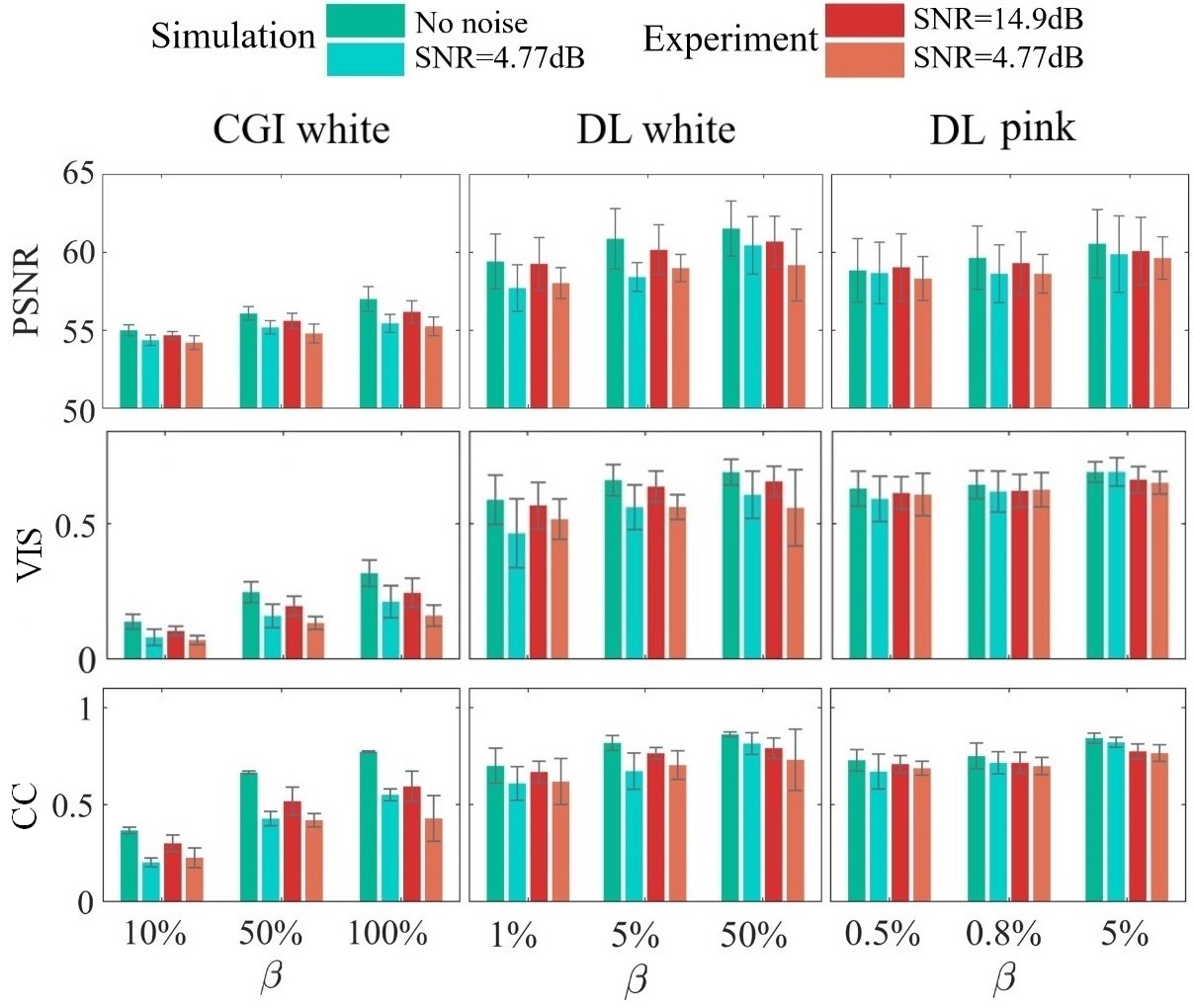}
\caption{PSNR, VIS, and CC for simulation and experiments of the block style digits 2, 3, 5, 6 with three categories: CGI white ($\beta$ of 10\%, 50\%, and 100\%), DL white ($\beta$ of 1\%, 5\%, and 50\%), and DL pink ($\beta$ of 0.5\%, 0.8\%, and 5\%).}
\label{fig:estimation}
\end{figure}

The results for all cases including simulation without and with noise, experiment with high and low SNR, are shown in Fig.~\ref{fig:estimation}. The $\mathrm{PSNR}$, $\mathrm{VIS}$, and $\mathrm{CC}$ all indicate that the CGIDL methods are much better than the traditional CGI method. Indeed, as shown in the simulation case, the image quality of CGIDL at 5\% is already better than CGI at full sampling rate for all the situations. When we compare the two DL methods, we see that in general DL pink is much better than DL white, as also suggested from Figs.~\ref{fig:simu_nonoise},~\ref{fig:simu_noise}, and~\ref{fig:exp_result}. Since the network is trained using $\mathrm{MSE}$ as the loss function, the $\mathrm{PSNR}$ of simulation without noise at 5\% is very similar for both cases. However, when the noise increases, the $\mathrm{PSNR}$ of DL white starts to decrease, while the $\mathrm{PSNR}$ of DL pink does not change much. The $\mathrm{VIS}$ and $\mathrm{CC}$ also have similar behavior as $\mathrm{PSNR}$. We note here that all three indicators suggest DL pink works better than the other two methods, in the experimental results with low SNR, DL pink of 5\% sampling rate is already better than the DL white with 50\% sampling rate.

\section{Conclusion}

In conclusion, we have demonstrated a deep-learning imaging method with pink noise patterns. The  DNN is trained using only simulation data from the handwriting dataset. The trained network can then be applied to various conditions, including objects outside the training set and experiments with strong noise. We have demonstrated imaging results with extremely low sampling rate both in simulation and experiments. We have also evaluated the quality of the images outside the training dataset for both simulation and experimental results, in terms of $\mathrm{PSNR}$, $\mathrm{VIS}$, and $\mathrm{CC}$. 

All results suggest that the DL pink scheme has a great advantage, especially in the low sampling region. This one-time, noise-robust, and non-experimental training CGIDL is eligible to be implemented in various situations and has a wide range of application prospects. The pink noise speckle patterns, trained DNN with various sampling rates, and their raw encoding programs are encapsulated and uploaded online \footnote{https://github.com/XJTU-TAMU-CGI/CGIDL}. People who need a quick sampling function on CGIDL can utilize this universal system to get ameliorated results in other CGIDL systems. Further works can reach to other imaging and spectroscopy systems by loss function adjustment and speckle pattern optimization, in order to get spatial, frequency, or time-resolution. In addition to results amelioration, DL may also have great potential to generate optimized speckle patterns for a variety of tasks.

\section*{Funding} Air Force Office of Scientific Research (Award No. FA9550-20-1-0366 DEF), Office of Naval Research (Award No. N00014-20-1-2184), Robert A. Welch Foundation (Grant No. A-1261), National Science Foundation (Grant No. PHY-2013771).

%\section{Acknowledgement.}
%H. Song and X. Nie contributed equally to this work.

%\noindent\textbf{Disclosures.} The authors declare no conflicts of interest.

\section*{Data availability.} The experimental and simulation data are available upon reasonable request. The trained DL networks and raw DL training and test codes are uploaded on the website: {\url{https://github.com/XJTU-TAMU-CGI/CGIDL}}.

\section*{Disclosures} The authors declare no conflicts of interest.

\end{document}